\documentclass[aps,prx,twocolumn,longbibliography,amsmath,amssymb,nofootinbib,longbibliography,10pt]{revtex4-1}
\usepackage{amsfonts,amssymb,amsmath}
\usepackage{graphics}
\usepackage{graphicx}
\usepackage{nicefrac}
\usepackage{mathrsfs}
\usepackage{enumerate}
\usepackage{mathtools}
\usepackage{textcomp}
\usepackage{verbatim}
\usepackage{bm}          
\usepackage{soul}
\usepackage{cancel}
\usepackage{upgreek}
\usepackage{txfonts}
\usepackage{color}
\usepackage[colorlinks=true, urlcolor=blue, linkcolor=blue]{hyperref}
\usepackage[papersize={8.5in,11in}]{geometry}
\usepackage{esint}
\newcommand{\be}{\begin{equation}}
\newcommand{\ee}{\end{equation}}

\renewcommand{\vec}[1]{\boldsymbol{#1}}

\def \v {{\bf v}}

\def \beq {\begin{eqnarray}}
\def \eeq {\end{eqnarray}}

\usepackage{makecell}

\begin{document}
\title{Transparent mirror effect in twist-angle-disordered bilayer graphene}
\author{Sandeep Joy, Saad Khalid, Brian Skinner}
\affiliation{Department of Physics, Ohio State University, Columbus, OH 43210, USA}
\date{\today}
\begin{abstract}
When light is incident on a medium with spatially disordered index of refraction, interference effects lead to near-perfect reflection when the number of dielectric interfaces is large, so that the medium becomes a ``transparent mirror."
We investigate the analog of this effect for electrons in twisted bilayer graphene (TBG), for which local fluctuations of the twist angle give rise to a spatially random Fermi velocity.
In a description that includes only spatial variation of Fermi velocity, we derive the incident-angle-dependent localization length for the case of quasi-one-dimensional disorder by mapping this problem onto one dimensional Anderson localization. The localization length diverges at normal incidence as a consequence of Klein tunneling, leading to a power-law decay of the transmission when averaged over incidence angle. In a minimal model of TBG, the modulation of twist angle also shifts the location of the Dirac cones in momentum space in a way that can be described by a random gauge field, and thus Klein tunneling is inexact. However, when the Dirac electron's incident momentum is large compared to these shifts, the primary effect of twist disorder is only to shift the incident angle associated with perfect transmission away from zero. 
These results suggest a mechanism for disorder-induced collimation, valley filtration, and energy filtration of Dirac electron beams, so that TBG offers a promising new platform for Dirac fermion optics.
\end{abstract}
\maketitle

Transmission of light through a medium with random refractive index is a well studied problem in optics \cite{wiersma_localization_1997, schwartz_transport_2007, PhysRevLett.100.013906}. Of particular interest is the problem of ``transparent mirrors", in which the comprising elements are all transparent, but when arranged in a many-layered stack they form a medium with near-perfect reflection. Even though the first theoretical treatment of a similar problem was proposed two centuries ago by Fresnel \cite{Tuckerman:s}, it was only relatively recently that Berry and Klein showed that the problem could be understood by analogy with Anderson localization \cite{Berry_1997}. The corresponding disorder-averaged transmission intensity decays exponentially with the number of dielectric interfaces.

The linearity of the energy dispersion for Dirac electrons offers an analogy with optics. This analogy has inspired previous authors to explore numerous analog optical phenomena in graphene, including but not limited to Veselago lensing \cite{Cheianov_2007}, sub-wavelength diffraction \cite{PhysRevLett.102.136803}, and the Goos-H{\"a}nchen effect \cite{PhysRevLett.102.146804}. For the same reason, the Dirac equation with velocity disorder strikingly resembles the problem of transmission of light through a medium with random refractive index \cite{PhysRevB.82.033413, PhysRevB.81.073407, Downing_2017}. The analogy with optics seemingly suggests that for a series of parallel one-dimensional domains of random Fermi velocity, one should expect exponential decay of the transmission coefficient. However, unlike in optics, Dirac electrons experience Klein tunneling, which implies that there is no reflection at normal incidence \cite{Katsnelson_2006}. This key difference suggests the idea of an angle-dependent localization length, which diverges as the incidence angle goes to zero. For an incident electron beam with a wide range of incident angles, the angle-averaged transmission may have a decay that is slower than exponential.

\begin{figure}[htb]
\centering
\includegraphics[width=1.0 \columnwidth]{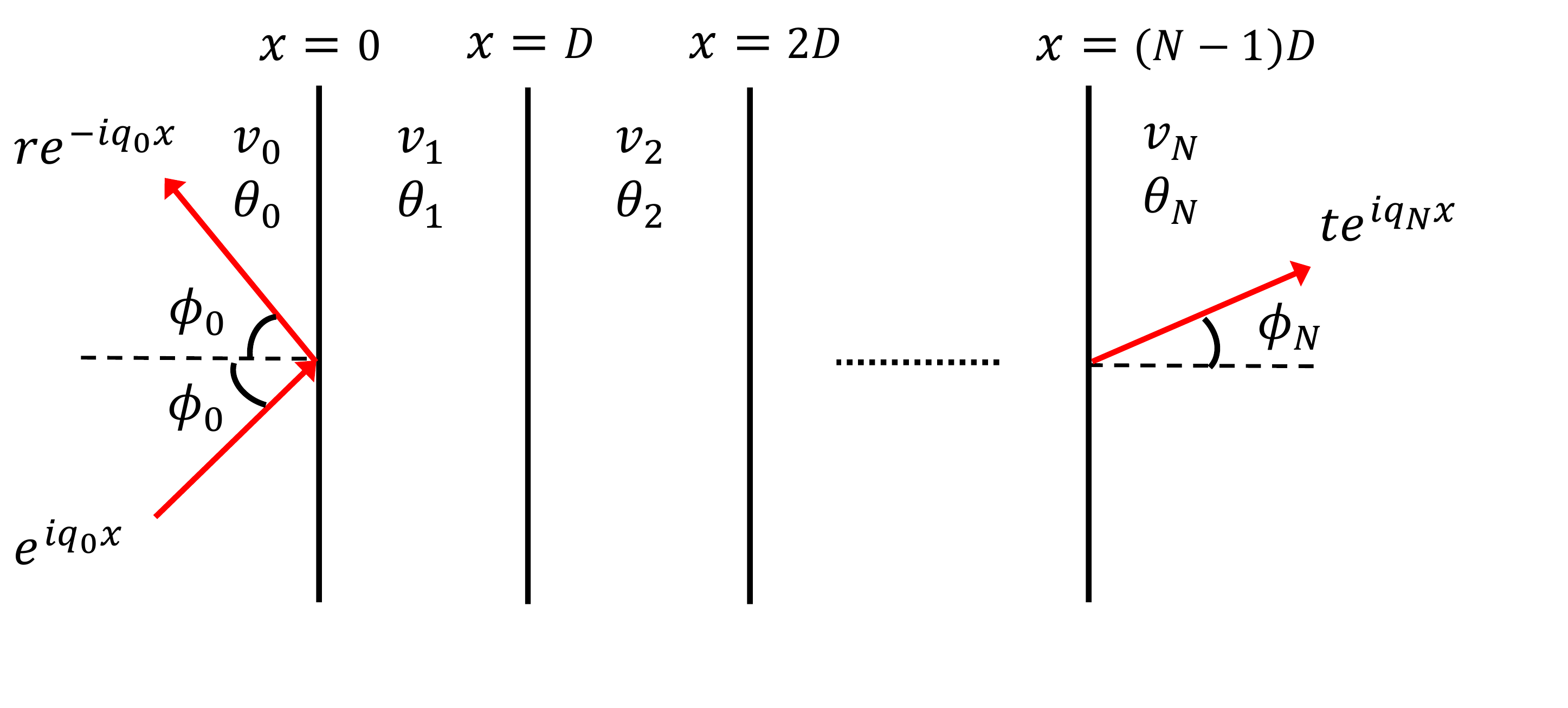}
\caption{Our model of quasi-one-dimensional  random domains of uniform thickness $D$. We have $N$ interfaces, such that each one is located at $x=jD$, where $j\in\left\{ 0,1,2,3,...,N-1\right\}$. The symbols $v_{j}$ and $\theta_{j}$ denote the Fermi velocity and twist angle, respectively, of the domain situated between $x=(j-1)D$ and $jD$. 
}
\label{fig:domains}
\end{figure}

Recent experiments mapping the spatial variation of twist angle in twisted bilayer graphene (TBG) provide a conspicuous example of a Dirac system with disordered velocity \cite{Uri_2020, kerelsky_maximized_2019, Yoo_2019, PhysRevResearch.2.023325, padhi2020transport}. Since the Fermi velocity in TBG depends sensitively on the twist angle \cite{Bistritzer_2011, PhysRevB.86.155449, PhysRevB.84.045436}, the formation of domains with different twist angle indicates a Fermi velocity disorder. While the effect of scalar and vector disorder potential on the transport of Dirac electrons has been explored theoretically \cite{PhysRevLett.100.156801, PhysRevB.76.195445, PhysRevLett.96.246802, Titov_2007, PhysRevB.99.014205, PhysRevB.85.104201}, the consequences of a spatially random Fermi velocity has not been addressed thoroughly. In this paper we study a model of electron transmission through a Dirac material with spatially modulated Dirac velocity. In particular, we work with the setup shown in Fig.~\ref{fig:domains}, where the velocity is modulated across parallel, quasi-one-dimensional domains with uniform thickness $D$. We assume step-like changes in velocity, which corresponds to the limit where $D$ is long compared to the electron wavelength and the thickness of the domain wall is small compared to the wavelength. Within this setup we consider two scenarios:
\begin{enumerate}[I.]
    \item The case of massless Dirac electrons for which only the Fermi velocity varies from domain to domain
    \item A minimal Hamiltonian for TBG, for which random domains of twist angle produce both a random velocity and a random gauge field
\end{enumerate}

It is worth mentioning that there are other methods to achieve Fermi velocity modulation in a Dirac material, in addition to variations in twist angle. Namely, one can use (i) strain engineering \cite{Naumis_2017, Amorim_2016} (ii) dielectric screening \cite{elias_dirac_2011, Hwang_2012, PhysRevB.81.073407} or (iii) a synthetic Dirac material with a super-lattice potential \cite{Park_2009, PhysRevB.79.241406, PhysRevLett.106.060503, Boada_2011}. Manipulating Fermi velocity via mechanical strain is a particularly well-studied method. Since strain is a tensor, in general it can change the velocity in anisotropic ways. Strain also shifts the location of the Dirac cone in momentum space so that Klein tunneling is not necessarily preserved, although it is preserved in the case of uniaxial strain in the zigzag direction \cite{PhysRevB.84.195404}. Dielectric screening by a substrate permits the modulation of Fermi velocity near the Dirac point via renormalization of electron-electron interactions. One can also make a synthetic Dirac dispersion by arranging atoms or creating an imposed potential energy with hexagonal symmetry.

\begin{figure}[htb]
\centering
\includegraphics[width=1.0 \columnwidth]{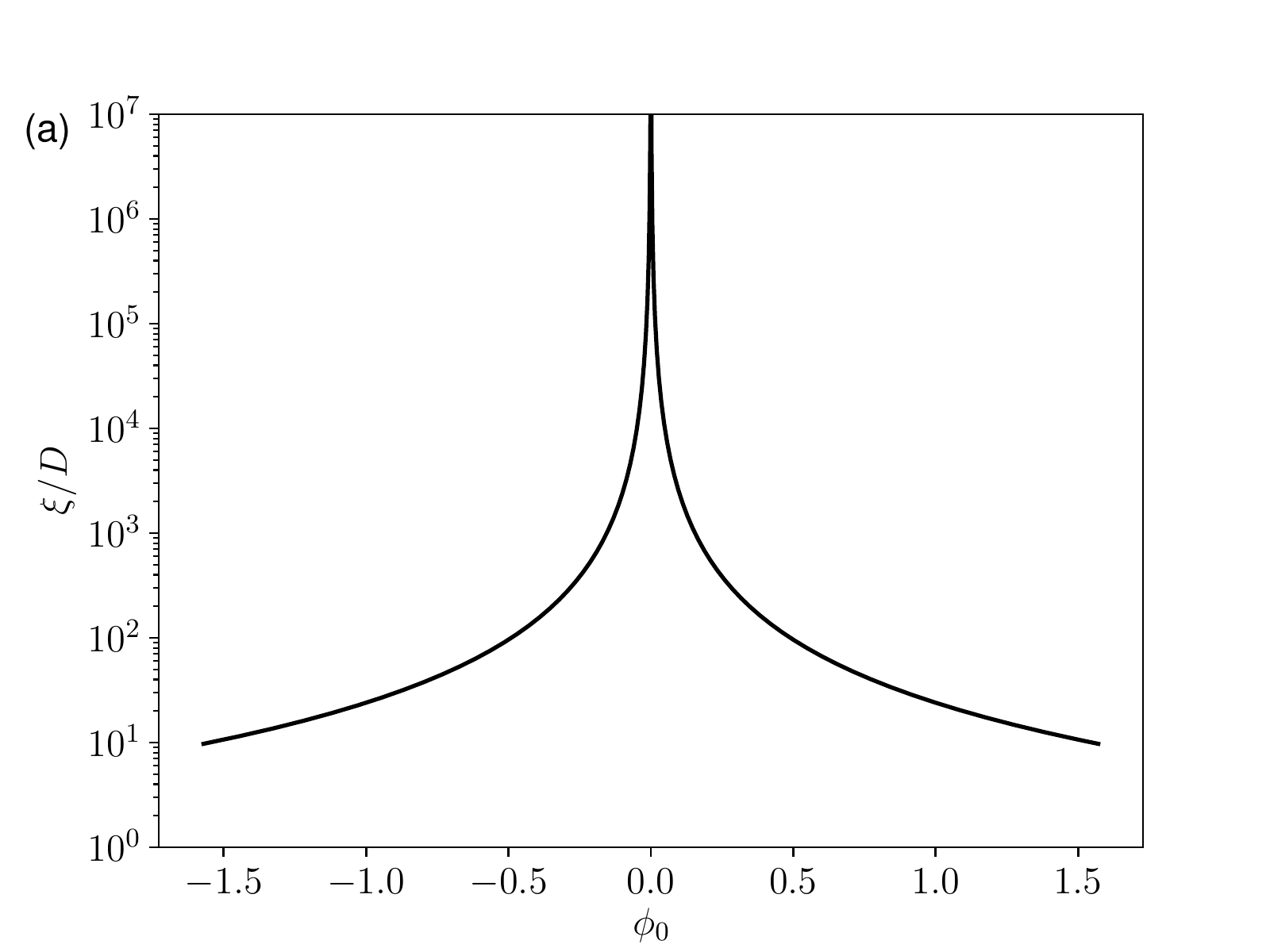}
\includegraphics[width=1.0 \columnwidth]{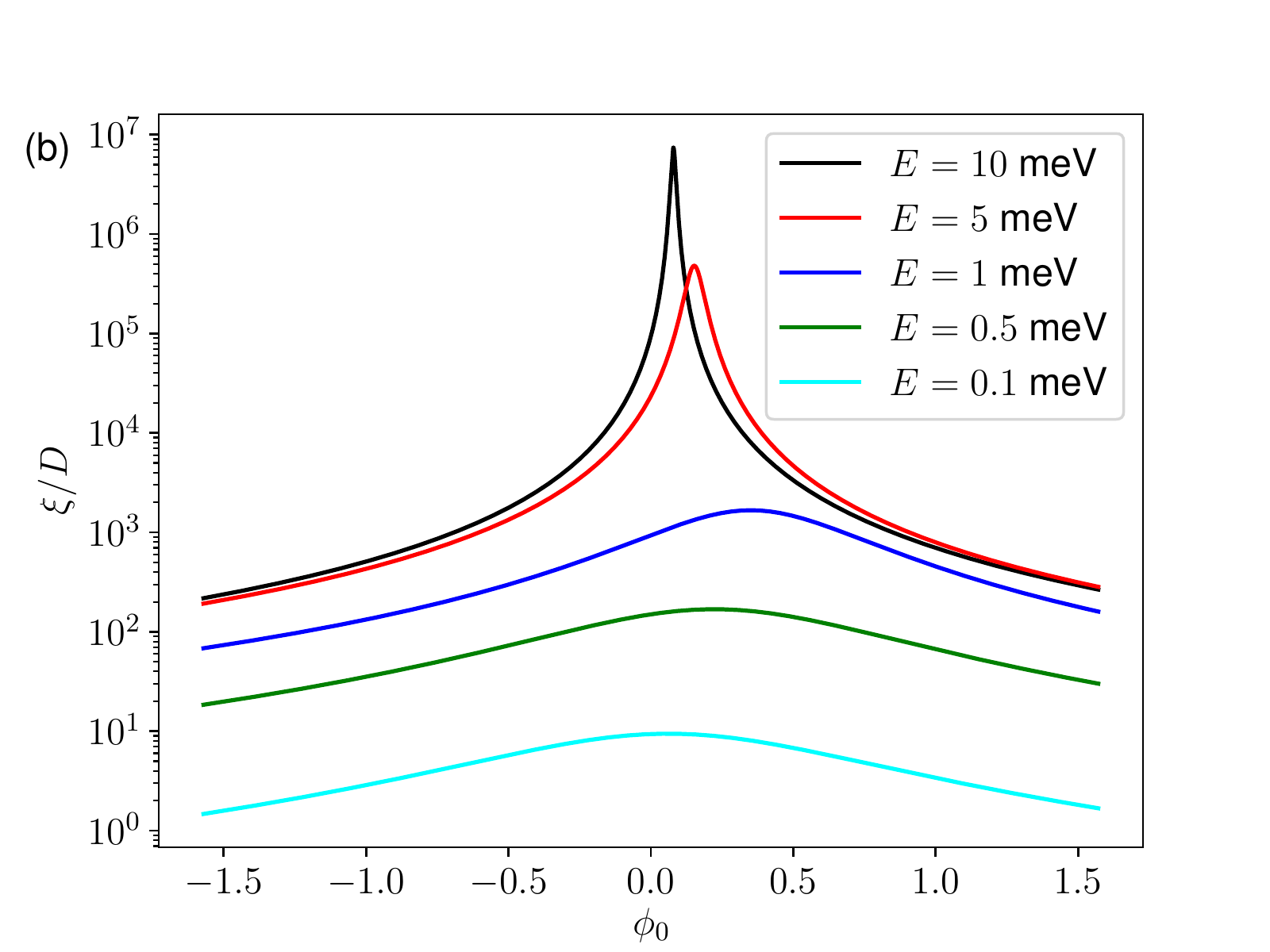}
\caption{The dependence of the localization length $\xi$ on the incident angle $\phi_0$. (a) For the case of massless Dirac electrons with Fermi velocity disorder only, the localization length diverges at normal incidence, $\phi_0 = 0$, irrespective of the electron energy. [See Eq.~(\ref{eq: xi})]
(b) For the case of a Hamiltonian describing local twist angle modulation in TBG, the incidence angle associated with maximal electron transmission is nonzero, and depends on the electron energy $E$ [see Eq.~(\ref{eq: TBGDWRi})]. This effect implies a disorder-induced collimation and energy filtration of electrons. The curves plotted here correspond to TBG near the magic angle, with 10\% variation in twist angle, as discussed in Sec.~\ref{sec:TBG}. 
}
\label{fig:mainresult}
\end{figure}

Our main results can be summarized as follows. In Sec.\ \ref{sec:MDF}, we show that the transmission of two-dimensional massless Dirac fermions through a medium with quasi one-dimensional random Fermi velocity can be exactly mapped onto the problem of conventional Anderson localization in one dimension. For small incident angle $\phi_{0}$, the localization length $\xi$ varies as 
$\xi \sim 1/\phi_0^2$ [see Fig.~\ref{fig:mainresult}(a)]. The diverging localization length at normal incidence is the manifestation of Klein tunneling in this system. The angle-averaged 
transmission in this system decays as 
$N^{-1/2}$, where $N$ is the number of interfaces. In Sec.\ \ref{sec:TBG}, twist angle disorder in TBG is mapped onto the problem of a Dirac Hamiltonian subjected to both velocity disorder and step changes of a magnetic vector potential. When the momentum of the incident electron is large compared to the shift in reciprocal space produced by the vector potential, Klein tunneling is recovered almost completely, although the incidence angle associated with perfect transmission is shifted away from zero [Fig.~\ref{fig:mainresult}(b)]. 
The resulting transmission similarly decays as $N^{-1/2}$ up to some very large $N$. 
In Sec.\ \ref{sec: condfano}, we relate our results for the electron transmission to the electrical conductivity and Fano factor. While we discuss our results everywhere in the context of graphene, it is worth mentioning that the results of Sec.\ \ref{sec:MDF} also hold true for any electronic system that exhibit properties which can be well described by the Dirac equation \cite{Vafek_2014}. 

\section{Massless Dirac Electrons with Random Fermi Velocity}
\label{sec:MDF}
In this section, we derive the transmission coefficient for Dirac electrons incident on the quasi-one-dimensional pattern of random velocity depicted in Fig.~\ref{fig:domains}. We assume, for the moment, that only the velocity is modulated as a function of the coordinate, so that electrons are described by a massless Dirac equation with position-dependent velocity. In Sec.\ \ref{sec:TBG}, we consider a more realistic model for twist-angle-disordered bilayer graphene, which includes the effects of the random gauge field that describes the shift of the Dirac point in momentum space.

The massless Dirac equation with position dependent velocity $v_F(\vec{r})$ is given by \cite{PhysRevB.81.073407}
\be
-\dot{\iota}\hbar\sqrt{v_{F}\left(\vec{r}\right)}\,\overrightarrow{\sigma}.\vec{\nabla}\left(\sqrt{v_{F}\left(\vec{r}\right)}\Psi\left(\vec{r}\right)\right)=E\Psi\left(\vec{r}\right),
\label{eq:DirEval}
\ee
where $\hbar$ is the reduced Planck constant, $\vec{\sigma}$ is the vector of Pauli matrices, and $E$ is the electron energy.  The wave function $\Psi\left(\vec{r}\right)$ is a two-component spinor
\be
\Psi\left(\vec{r}\right)=\left[\begin{array}{c}
\psi_{A}\left(\vec{r}\right)\\
\psi_{B}\left(\vec{r}\right)
\end{array}\right].
\label{spinor}
\ee
If we define the auxillary spinor $\widetilde{\Psi}\left(\vec{r}\right)\equiv\sqrt{v_{F}\left(\vec{r}\right)}\Psi\left(\vec{r}\right)$, the Dirac equation becomes
\be
-\dot{\iota}\hbar v_{F}\left(\vec{r}\right)\overrightarrow{\sigma}.\vec{\nabla}\widetilde{\Psi}\left(\vec{r}\right)=E\widetilde{\Psi}\left(\vec{r}\right).
\label{eq: DirEvalNew}
\ee
We consider the case where the Fermi velocity varies only in the $x$-direction, i.e.\ we assume $v_{F}\left(\vec{r}\right)=v_{F}\left(x\right)$. The translational invariance along the $y$-axis allows us to write the following ansatz
\be
\widetilde{\Psi}\left(\vec{r}\right)=\left[\begin{array}{c}
\widetilde{\psi}_{A}\left(x\right)\\
\widetilde{\psi}_{B}\left(x\right)
\end{array}\right]\times e^{\dot{\iota}q_{y}y}.
\ee
Substituting this ansatz into Eq.\ (\ref{eq: DirEvalNew}), we get the following coupled differential equations
\begin{align}
\begin{split}
 -\dot{\iota}\hbar v_{F}\left(x\right)\left(\partial_{x}+q_{y}\right)\widetilde{\psi}_{B}\left(x\right)&=E\widetilde{\psi}_{A}\left(x\right),  \\-\dot{\iota}\hbar v_{F}\left(x\right)\left(\partial_{x}-q_{y}\right)\widetilde{\psi}_{A}\left(x\right)&=E\widetilde{\psi}_{B}\left(x\right).
 \end{split}
\end{align}
We are interested in a situation where the Fermi velocity is piece-wise constant; let it be $v_{F}$ locally. In this situation, we can manipulate the above coupled first order differential equation to get the following uncoupled second order differential equation
\be
\left[\frac{d^{2}}{dx^{2}}+\left(\left(\frac{E}{\hbar v_{F}}\right)^{2}-q_{y}^{2}\right)\right]\widetilde{\psi}_{A}\left(x\right)=0.
\label{eq: cde}
\ee
The solution to this equation is a linear superposition of plane waves moving in the left and right directions:
\be
\widetilde{\psi}_{A}\left(x\right)=ae^{\dot{\iota}qx}+be^{-\dot{\iota}qx},
\ee
where $q\equiv\sqrt{\left(\frac{E}{\hbar v_{F}}\right)^{2}-q_{y}^{2}}$ is the wavevector along the $x$-direction. The angle $\phi$ between the $x$ and $y$ momenta of the electron satisfies
\be
\sin\phi = \frac{\hbar v_{F}q_{y}}{E}.
\ee
Plugging $\widetilde{\psi}_{A}$ into Eq.~(\ref{eq: cde}) and using the definition of $\phi$ we can write $\widetilde{\psi}_{B}$ as
\be
\widetilde{\psi}_{B}=ae^{\dot{\iota}\left(qx+\phi\right)}-be^{-\dot{\iota}\left(qx+\phi\right)}.
\ee
Now we can write the whole Dirac auxiliary spinor $\widetilde{\Psi}$ as
\be
\widetilde{\Psi}=\mathcal{N}\left(ae^{\dot{\iota}qx}\left[\begin{array}{c}
1\\
e^{\dot{\iota}\phi}
\end{array}\right]+be^{-\dot{\iota}qx}\left[\begin{array}{c}
1\\
-e^{-\dot{\iota}\phi}
\end{array}\right]\right)e^{\dot{\iota} q_{y}y}.
\ee
We have added an overall normalization factor of $\mathcal{N}$. In this paper the normalization is chosen such that $j_{x}=v_{F}\Psi^{\dagger}\sigma_{X}\Psi$, which is the probability current of Dirac electrons along $x$-direction, is normalized to $\left|a\right|^{2}-\left|b\right|^{2}$. This choice is $\mathcal{N}=\nicefrac{1}{\sqrt{2\cos{\phi}}}$.

\begin{figure}[htb]
\centering
\includegraphics[width=1.0 \columnwidth]{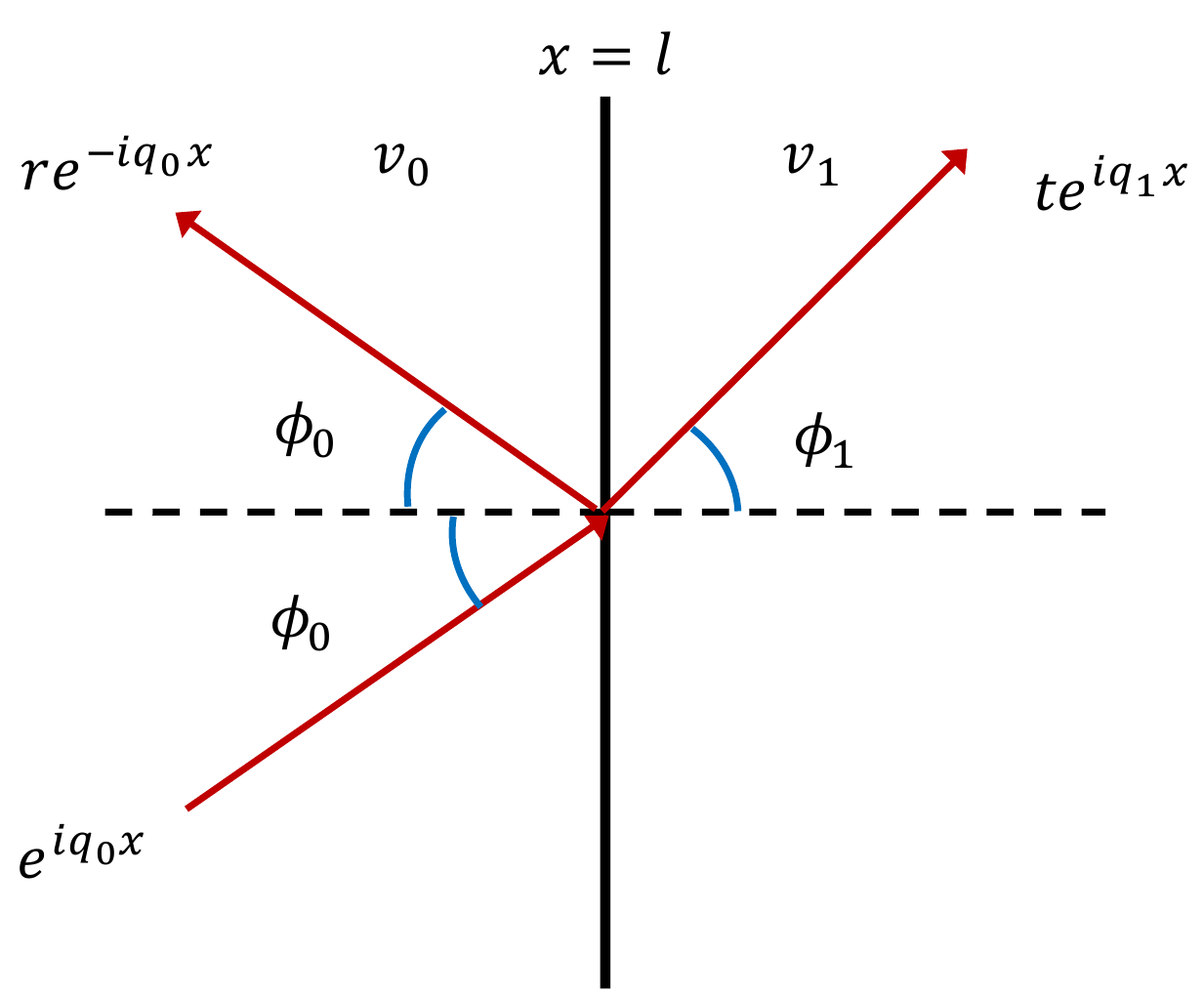}
\caption{Scattering of a Dirac electron at step velocity barrier located at $x=l$. The velocity changes abruptly at the interface from $v_{0}$ to $v_{1}$. The Dirac electron undergoes both reflection and refraction, as in the analogous optical problem.}
\label{fig:single}
\end{figure}

Before considering the transmission through an arbitrary number of interfaces, we first examine the case of a single interface [see Fig.\ \ref{fig:single}] at an arbitrary position $x = l$. 
The Fermi velocity across the interface is given by
\be
v_{F}\left(x\right)=v_{0}\,\Theta\left(l-x\right)+v_{1}\,\Theta\left(x-l\right),
\label{eq: Snell}
\ee
where $\Theta\left(x\right)$ is the Heaviside step function. Using the conservation of $y$-momentum across the interface we obtain the analog of Snell's law:
\be
\sin\phi_{1}=
\frac{v_{1}}{v_{0}} \sin\phi_{0}.
\ee
If $v_{1} > v_{0}$, then there exists a critical angle $\phi_c = \arcsin(v_0/v_1)$ such that $\phi > \phi_c$ corresponds to total internal reflection. At such large angles the electron wave function crosses the barrier only through an evanescent mode, which decays exponentially in amplitude with $x - l$. In this paper we work in the limit where the domain width $D$ is sufficiently large that these evanescent modes can be neglected (In the numerical results presented below and in Fig. \ref{fig:mainresult}, we set $(E/\hbar v_0)D = 4 \pi$.). 

The boundary condition for the Dirac Hamiltonian at the interface is given by the continuity of both components of the auxiliary spinor $\widetilde{\Psi}$. We therefore arrive at the transfer matrix $M$ which relates the amplitudes $a_0$, $b_0$ on the left region to the amplitudes $a_1$, $b_1$ of the right region:
\be
\left[\begin{array}{c}
a_{1}\\
b_{1}
\end{array}\right]=M\left[\begin{array}{c}
a_{0}\\
b_{0}
\end{array}\right],
\ee
where
\be
M=\left[\begin{array}{cc}
\alpha & \beta\\
\beta^{*} & \alpha^{*}
\end{array}\right].
\label{eq: umpum}
\ee
The complex numbers $\alpha$ and $\beta$ are given by
\begin{align}
    \alpha&=\frac{e^{\dot{\iota}\left(q_{0}-q_{1}\right)l}\left(e^{\dot{\iota}\phi_{0}}+e^{-\dot{\iota}\phi_{1}}\right)}{2\sqrt{\cos\left(\phi_{0}\right)\cos\left(\phi_{1}\right)}},\\\beta&=\frac{e^{-\dot{\iota}\left(q_{0}+q_{1}\right)l}\left(e^{-\dot{\iota}\phi_{1}}-e^{-\dot{\iota}\phi_{0}}\right)}{2\sqrt{\cos\left(\phi_{0}\right)\cos\left(\phi_{1}\right)}}.    
\end{align}
We can calculate the transmission probability across a single interface by defining the reflection and transmission amplitudes, $r$ and $t$, respectively [see Fig.\ \ref{fig:single}], such that
\be
\left[\begin{array}{c}
t\\
0
\end{array}\right]=\left[\begin{array}{cc}
M_{11} & M_{12}\\
M_{21} & M_{22}
\end{array}\right]\left[\begin{array}{c}
1\\
r
\end{array}\right].
\ee
The corresponding transmission probability $T_{1}=1-\left|r\right|^{2}$ is
\be
T_{1}=\frac{1}{\left|M_{22}\right|^{2}}=\frac{2\cos\left(\phi_{0}\right)\cos\left(\phi_{1}\right)}{1+\cos\left(\phi_{0}+\phi_{1}\right)}.
\label{eq: T1}
\ee
With the help of Snell's law the transmission probability can be rewritten as
\be
T_{1}=\frac{2\cos\phi_{0}\sqrt{1-\left(u_{1}\sin\phi_{0}\right)^{2}}}{1+\cos\phi_{0}\sqrt{1-\left(u_{1}\sin\phi_{0}\right)^{2}}-u_{1}\left(\sin\phi_{0}\right)^{2}},
\ee
where $u_{1}= v_1 / v_0$. Under normal incidence, $\phi_0=0$, we have $T_{1}=1$, which is the consequence of Klein tunneling. This expression also gives $T_1 = 1$ when $u_1 = 1$, which is the condition of perfect transmission when there is no barrier at all.

When there are multiple interfaces, the transfer matrices for each interface follow the composition rule, such that for $N$ interfaces 
\begin{align}
    \left[\begin{array}{c}
a_{N} \nonumber \\
b_{N}
\end{array}\right]&=M_{N}\left[\begin{array}{c}
a_{N-1}\\
b_{N-1}
\end{array}\right]\\&=M_{N}...M_{2}M_{1}\left[\begin{array}{c}
a_{0}\\
b_{0}
\end{array}\right].
\end{align}
We can use this idea to calculate the transmission coefficient when there are two interfaces present. In this case we have three consecutive regions with velocities $v_{0}$, $v_{1}$, and $v_{2}$ respectively. The transmission amplitude $t_{12}$ across the two interfaces combined can be found to be 
\be
t_{12}=\frac{t_{1}t_{2}}{1-r_{1}'r_{2}},
\label{eq: cohsum1}
\ee
which is equivalent to the following geometric series
\be
t_{12}=t_{2}t_{1}+t_{2}r_{1}'r_{2}t_{1}+t_{2}r_{1}'r_{2}r_{1}'r_{2}t_{1}+... \ .
\label{eq: cohsum2}
\ee
Here, $r_{1}'$ is the reflection amplitude for the waves incident from the right on the first interface if it was the only one present. Equation (\ref{eq: cohsum2}) provides a simple geometric interpretation of Eq.\ (\ref{eq: cohsum1}): when we combine two interfaces the net transmission amplitude is the coherent sum of all the multiply reflected waves in the region between the two interfaces \cite{Berry_1997}. The net transmission coefficient, $\mathcal{T}_{2}$, across both interfaces is
\be
\mathcal{T}_{2}=\frac{T_{1}T_{2}}{\left|1+\sqrt{R_{1}R_{2}}e^{\dot{\iota}\Phi_{12}}\right|^{2}},
\label{eq: T12}
\ee
where $T_{2}$ and $R_{2}$ represent the transmission and reflection probability, respectively if only the second interface was present. Here, $\Phi_{12} = 2q_{1}D$ is the total phase accumulated during one complete internal reflection in the sandwiched region between the two interfaces. 

\begin{figure}[htb]
\centering
\includegraphics[width=1.0 \columnwidth]{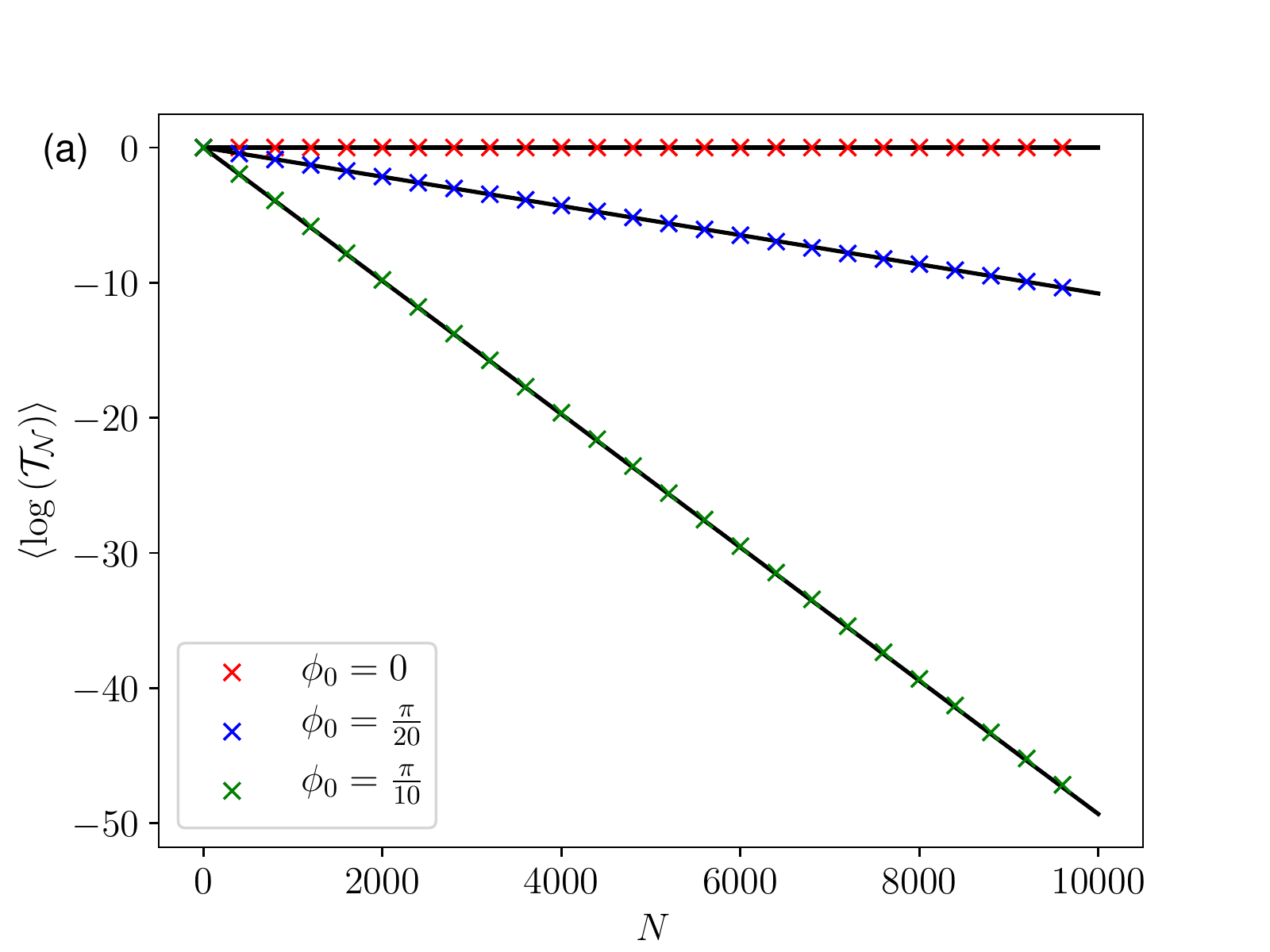}
\includegraphics[width=1.0 \columnwidth]{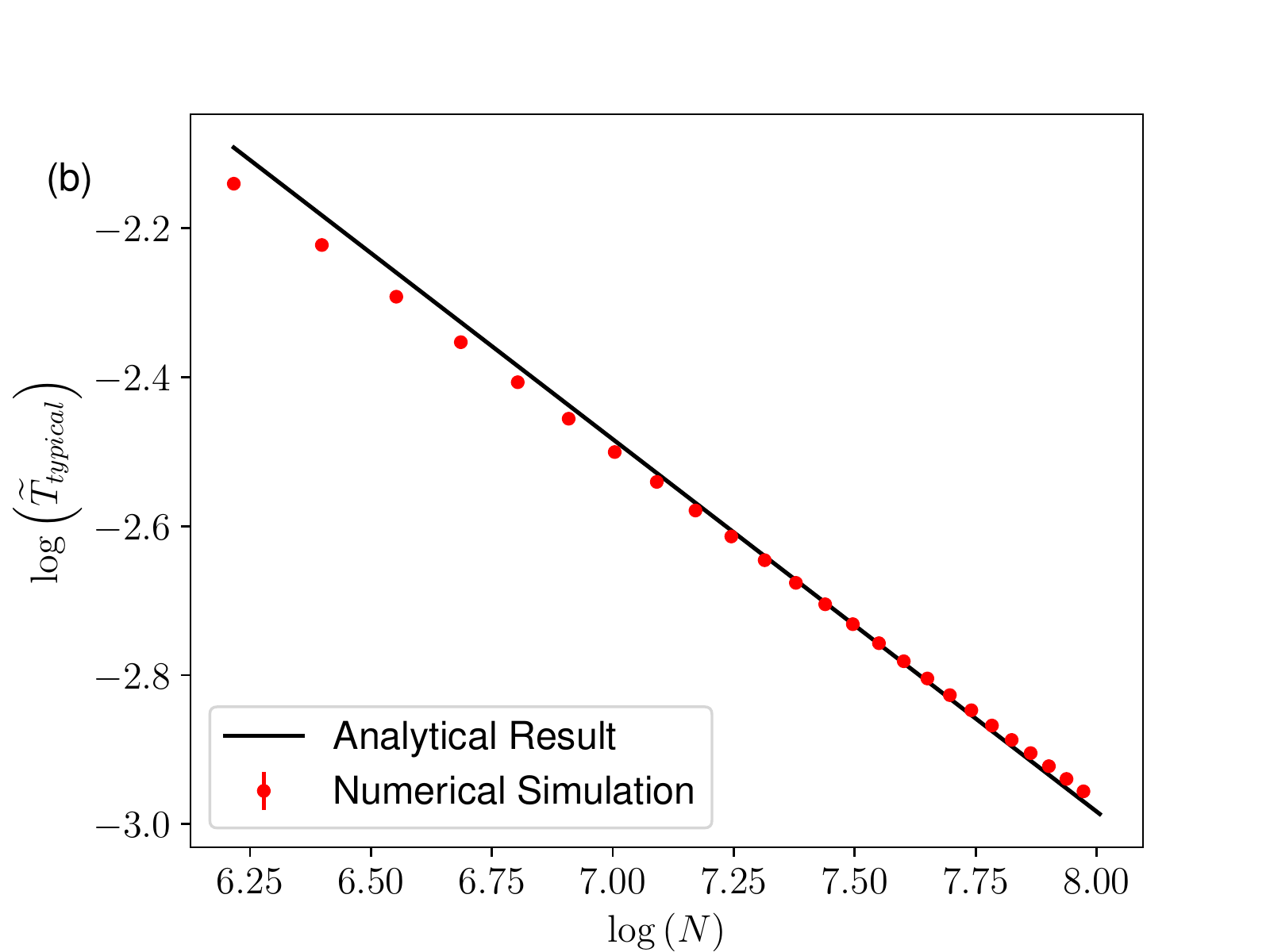}
\caption{Decay of the transmission amplitude with the number of interfaces for Dirac fermions with only velocity disorder. (a) The log-averaged transmission, $\left\langle \log\left(\mathcal{T}_{N}\right)\right\rangle$ is plotted as a function of the number of interfaces, $N$, for different incident angles $\phi_{0}=0, \pi/20$, and $\pi/10$, with the range of relative velocity disorder $w = 1$. Crosses denote results from a numerical simulation, and analytical results are plotted as solid lines, using Eq.~(\ref{eq: logTN}).
(b) The logarithm of angle-averaged typical transmission $\log\left(\widetilde{T}_{\text{typical}}\right)$ is plotted as a function of $\log\left(N\right)$, for both our numeric simulation (symbols) and the analytical result of Eq.\ (\ref{eq: angavgttyp}) (line).}
\label{fig:avgvarlogT}
\end{figure}

Let us calculate the average of the logarithm of the transmission coefficient across two interfaces. (Throughout this paper, we use ``log'' to denote the natural logarithm.) As we explain below, the transmission $\mathcal{T}_N$ is not self-averaging in the limit of large $N$, and instead we deal with $\log\left(\mathcal{T}_{N}\right)$, which is self-averaging. The average is taken over the values of the random velocities,  $u_{1}= v_1 / v_0$ and $u_{2}= v_2 / v_0$. We assume the following box distribution for the variables $u_{i}$:
\be
P\left[u_{i}\right]=\left\{ \begin{array}{c}
1/w\qquad\text{if}\qquad1-\frac{w}{2}\le u\le1+\frac{w}{2}\\
0\qquad\text{otherwise}
\end{array}\right..
\label{eq: PDF}
\ee
Taking the logarithm of both sides of Eq.\ (\ref{eq: T12}) and averaging over the disorder yields
\begin{align}
    \begin{split}
    \left\langle \log\left(\mathcal{T}_{2}\right)\right\rangle =&\left\langle \log\left(T_{1}\right)\right\rangle +\left\langle \log\left(T_{2}\right)\right\rangle \\&-2\left\langle \log\left(\left|1+\sqrt{R_{1}R_{2}}e^{\dot{\iota}\Phi_{12}}\right|\right)\right\rangle, 
    \end{split}
    \label{eq: logT12}
\end{align}
where $\left\langle. \right\rangle$ denotes the disorder average. The phase, $\Phi_{12}$, depends very sensitively on the value of $u_{1}$ when $D$ is large compared to the electron wavelength, which is the case we are considering. On the other hand, $R_{1}$ and $R_{2}$ vary slowly with the velocity. Consequently, the term $e^{\dot{\iota}\Phi_{12}}$ oscillates rapidly with $u_1$ and averages to zero as we integrate over different values of the velocity, giving $\left\langle \log\left(\mathcal{T}_{2}\right)\right\rangle \simeq\left\langle \log T_{1}\right\rangle +\left\langle \log T_{2}\right\rangle$. Equation (\ref{eq: logT12}) directly implies that the disorder-averaged logarithm of the transmission is additive when multiple interfaces are present. Generalization of this equation to many interfaces is now straightforward, since in the log-average the interfaces are effectively decoupled:
\begin{align}
        \left\langle \log\left(\mathcal{T}_{N}\right)\right\rangle &=\sum_{i=1}^{N}\left\langle \log T_{i}\right\rangle \nonumber \\  &\approx N\left\langle \log T_{i}\right\rangle , 
    \label{eq: logTN}
\end{align}
where
\be
T_{i}=\frac{2\sqrt{1-u_{i}^{2}\sin^{2}\phi_{0}}\sqrt{1-u_{i-1}^{2}\sin^{2}{\phi}_{0}}}{1+\sqrt{1-u_{i}^{2}\sin^{2}\phi_{0}}\sqrt{1-u_{i-1}^{2}\sin^{2}\phi_{0}}-u_{i}u_{i-1}\sin^{2}\phi_{0}}.
\ee
In the second line of Eq.\ (\ref{eq: logTN}), there is an approximate sign because for the very first term $\log T_{1}$, averaging is done only over the value of $u_{1}$. However, for every other interface $i > 1$, the averaging of $\log T_{i}$ is performed over both $u_{i-1}$ and $u_{i}$. Equation (\ref{eq: logTN}) demonstrates that the average logarithm of the transmission is a self-averaging quantity. The exponential of the log-averaged transmission gives the typical transmission, which corresponds to the median outcome in an ensemble average:
\be
T_{\text{typical}}\equiv \, e^{\left\{ \left\langle \log\left(\mathcal{T}_{N}\right)\right\rangle \right\} }=e^{-N D/\xi}.
\ee
Here we have defined the localization length $\xi$ such that $D/\xi=-\left\langle \log\left(T_{i}\right)\right\rangle$. The exact expression for the inverse localization length is given by
\be
\frac{D}{\xi}=-\frac{1}{w^{2}}\int_{1-\frac{w}{2}}^{1+\frac{w}{2}}\int_{1-\frac{w}{2}}^{1+\frac{w}{2}}du_{i}du_{i-1}\log T_{i}.
\ee
Defining the reflection coefficient for the $i$th interface as $R_i = 1 - T_i$  and Taylor expanding for small $\phi_0$ gives
\be
R_{i}\approx\frac{\phi_{0}^{2}}{4}\left(u_{i}-u_{i-1}\right)^{2}.
\ee
This result implies that
\be
\frac{D}{\xi} \approx \frac{w^{2}\phi_{0}^{2}}{24}.
\label{eq: xi}
\ee
When $\phi_{0}\rightarrow0$, the localization length diverges, as mentioned above. We confirm this expression for $\xi$ using a numerical simulation of the transfer matrix.  The result is plotted in Fig.\  \ref{fig:avgvarlogT}(a) [see also Fig.\ \ref{fig:mainresult}(a)], and is compared with the exact result derived in Eq.~(\ref{eq: logTN}).

When electrons are incident with a range of different $y$-momenta, the more physically-relevant quantity is the initial-angle-averaged typical transmission,
\be
\widetilde{T}_{\text{typical}}\equiv\frac{1}{\pi}\int_{-\phi_{c}}^{\phi_{c}}d\phi_{0}\;{T}_{\text{typical}}\left(\phi_{0}\right).
\label{eq: typicaldef}
\ee
This quantity corresponds to the average intensity of Dirac electrons transmitted across a sample, given an a initial wave-packet of varying initial angle of incidence.
As we discuss in Sec.\ \ref{sec: condfano}, this angle-averaged transmission is relevant for determining the electrical conductance. For simplicity we have assumed a uniform distribution of the initial angle. For large $N$, the leading order behavior of $\widetilde{T}_{\text{typical}}$ can be found by extending the limits to infinity, which gives
\be
\widetilde{T}_{\text{typical}} = \sqrt{\frac{24}{\pi w^{2}}} \frac{1}{\sqrt{N}}.
\label{eq: angavgttyp}
\ee
This equation constitutes one of the central results of our paper. While, for a given incident angle, the mapping to Anderson localization implies an exponential decrease of the transmission with distance, the divergence of the localization length for small angles gives an angle-averaged typical transmission that decays only as a slow power law, $N^{-1/2}$. This dependence is verified by a numerical simulation in Fig.\ \ref{fig:avgvarlogT}(b). One implication of this result is that, for an incoming electron beam with a wide dispersion of incident angles, only a small fraction 
$\sqrt{\frac{24}{\pi w^2 N}}$ is transmitted to a depth $N$.  These transmitted electrons are narrowly collimated to have near-normal transmission, and this collimation becomes increasingly tight as $N$ is increased.


The statistical properties of the transmission probability can be understood by noting that, for a fixed incident angle $\phi_0$, there is an exact analogy between our problem and the problem of Anderson localization in a disordered $1$D wire. The resulting distribution of $\mathcal{T}_{N}$ is log-normal for sufficiently large $N$ \cite{mller2010disorder}
\be
P\left(\log\mathcal{T}_{N}\right)=\frac{1}{2\sqrt{\pi N\left\langle R_{1}\right\rangle }}\exp{\left(-\frac{\left(\log\mathcal{T}_{N}+N\left\langle R_{1}\right\rangle \right)^{2}}{4N\left\langle R_{1}\right\rangle }\right)}.
\label{eq: lognormal} 
\ee
It is also possible to derive a more general differential equation that describes $P\left(\mathcal{T}_{N}\right)$ at not-too-large $N$. The corresponding calculations are discussed in Appendix \ref{sec: fokk} and the references therein.

\section{Twist angle domains in a model of TBG}
\label{sec:TBG}

\begin{figure}[htb]
\centering
\includegraphics[width=1.0 \columnwidth]{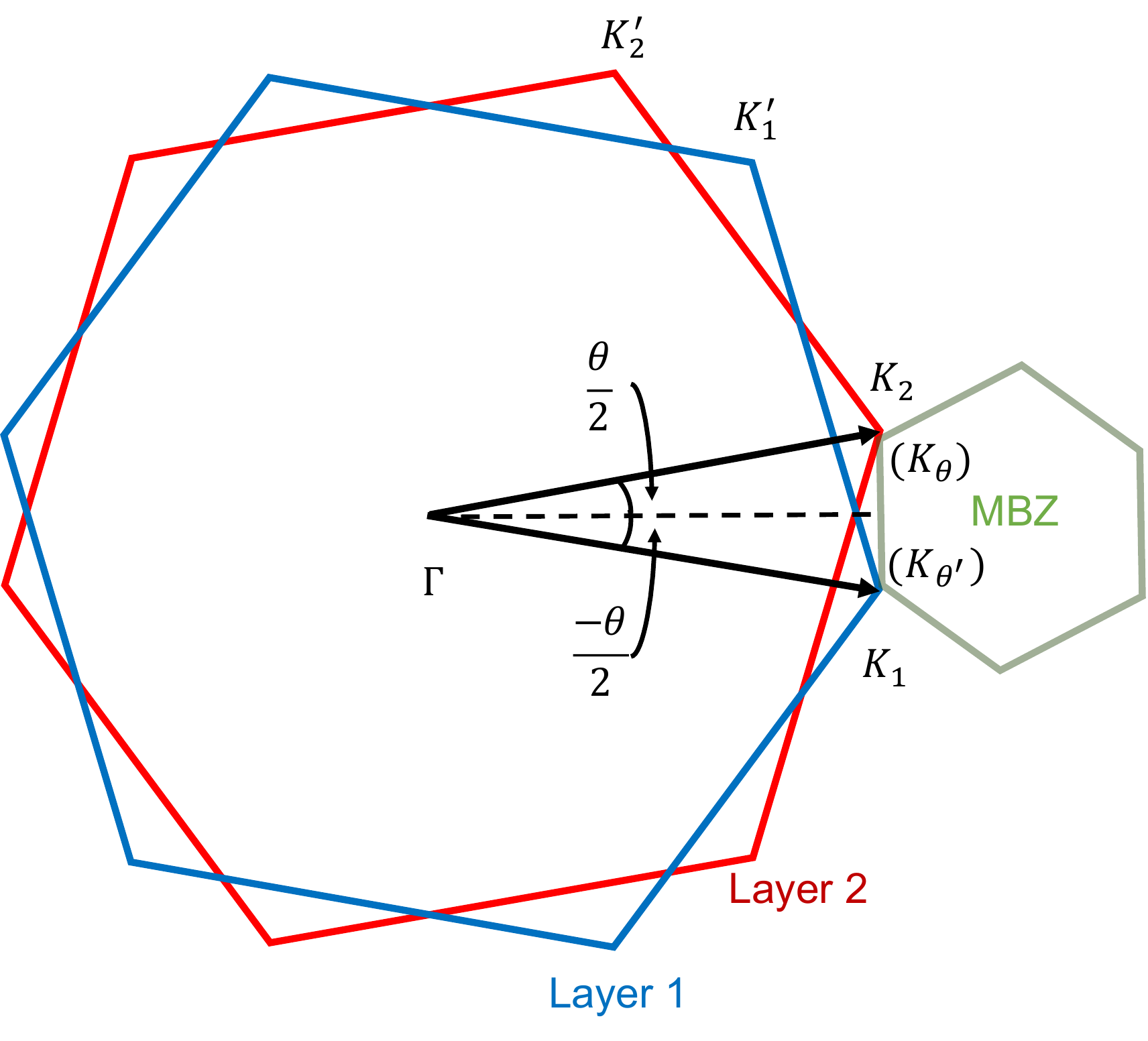}
\caption{A schematic illustration of the moir\'{e} Brillouin zone (MBZ) of twisted bilayer graphene (green hexagon); the horizontal and vertical coordinates in the diagram correspond to $p_x$ and $p_y$, respectively. The red and blue hexagons represent the Brillouin zone of the lower and upper layer of bare graphene, respectively. Here $\theta$ is the global twist angle between the two layers. $\Gamma$ corresponds to the laboratory reference frame, while $K_{1,2}$ and $K'_{1,2}$ denote the locations of the two Dirac points for layers $1, 2$. 
}
\label{fig:mbz}
\end{figure}

Twisted bilayer graphene consists of stacked sheets of graphene with a rotational mismatch between the individual layers. Twisting creates a moir\'{e} superlattice of periodicity, which for small twist angles can be more than two orders of magnitude larger than the lattice constant of monolayer graphene.  The corresponding moir\'{e} brillouin zone (MBZ) is relatively small. The MBZ is depicted schematically in Fig.\ \ref{fig:mbz}, along with the brillouin zones (BZs) of individual graphene layers. Coupling between the two twisted layers leads to a hybridized dispersion relation with a renormalized Fermi velocity, which for certain ``magic angles'' can be as dramatic as reducing $v_F$ to a small percentage of its bare value  \cite{Bistritzer_2011, PhysRevB.86.155449, PhysRevX.8.031087}. Near these magic angles the Fermi velocity depends sensitively on the twist angle, so that even a mild twist angle disorder corresponds to a large velocity disorder. The twist angle also changes the size of the MBZ and shifts the Dirac points in momentum space, which is evident from Fig.\ \ref{fig:mbz}. As explained by Padhi et al.\ \cite{padhi2020transport}, when the Dirac points shift in momentum space across a domain wall it eliminates the possibility of Klein tunneling for low-energy electrons. Reference \onlinecite{padhi2020transport} focuses on this effect and neglects the change in velocity across the domain wall.  

In fact, the question of transmission across a twist angle domain wall is crucially determined by the details of how the twist angle is altered. For example, if one of the two layers is fixed while the other is free to twist, then one of the two Dirac points of the MBZ also remains fixed in momentum space. (As illustrated in Fig.~\ref{fig:mbz}, only one of the two larger BZs is rotated in this case, so that only one of the two Dirac points in the MBZ is shifted.) Correspondingly, this situation produces Klein tunneling at one of the two nodes of the MBZ. If this situation persists for every twist angle domain (i.e., the same is layer always twisting), then the results from the previous section remain valid for one of the two valleys, including the diverging localization length at $\phi_{0} = 0$. On the other hand, the Dirac point corresponding to the other valley wanders from one region of the sample to another, and, as we show below, the maximum of transmission for this valley occurs at a different, nonzero angle. This situation implies a disorder-induced valley filtration in which electrons from only one of the two valleys are transmitted at normal incidence. 

A more general scenario would involve both layers twisting from one domain to another, with a well-defined mean global twist angle. In the remainder of this section we assume that all twist angle changes happen symmetrically (as in Ref.\ \onlinecite{padhi2020transport}), and we describe the Hamiltonian for TBG reported in Ref.\ \onlinecite{PhysRevB.84.045436}.

Let $v_{F}$ be the Fermi velocity of pristine graphene, and $K$ be the difference in momentum between its two Dirac points. Let us define the mean global twist angle of the TBG system to be $\theta$, and we will denote the mean global position of the $K$ point of the MBZ as $\vec{K}_{\theta}$ (and similarly the $K'$ point of the MBZ as $\vec{K}_{\theta}^{'}$). The length of the MBZ is given by $\left|\Delta\vec{K_{\theta}}\right|=2K\sin\left(\frac{\theta}{2}\right)$, which corresponds to the separation between $\vec{K}_{\theta}$, and $\vec{K}_{\theta}^{'}$ . Using the continuum description \cite{PhysRevB.84.045436}, the renormalized Fermi velocity is $v_{0}=2\hbar v_{F}^{2}\left|\vec{K_{\theta}}\right|/\tilde{t}_{\perp}$, where $\tilde{t}_{\perp}$ is the interlayer coupling strength between the individual layers. 

Now consider an arbitrary domain with twist angle $\theta+\epsilon$, where $\epsilon$ is the fluctuating part of the twist angle which can be either positive or negative. To leading order in $\epsilon$, the $\vec{K}_{\theta+\epsilon}$ point of the MBZ in the new region is located at $\vec{K}_{\theta}+\vec{\delta}$, where
\be
    \vec{\delta}=\frac{K\epsilon}{2}\left(-\sin\left(\frac{\theta}{2}\right),\cos\left(\frac{\theta}{2}\right)\right).
\ee
The new Fermi velocity is found to be
\be
       v \simeq v_{0}\left(1+\frac{\epsilon\cot\frac{\theta}{2}}{2}\right).
       \label{eq: newvf}
\ee
The Hamiltonian in the vicinity of this Dirac point is given by
\be
\widehat{H}=v\vec{\sigma} \cdot \left(\widehat{\vec{P}}-\vec{K_{\theta}}-\vec{\delta}\right),
\ee
where $\widehat{\vec{P}} = (\widehat{p}_x, \widehat{p}_y)$, and $\widehat{p}_{x}$ and $\widehat{p}_{y}$ are the momenta operators defined with respect to the $\Gamma$ point of the monolayer graphene (which corresponds to the lab frame of reference, and is fixed regardless of twist angle, see Fig.\ \ref{fig:mbz}). 

Let us define new momenta operators $\widehat{q}_{x}$ and $\widehat{q}_{y}$, which measure the momenta with respect to the mean global $K$ point of the TBG
\begin{align}
    \widehat{q}_{x}&=-\dot{\iota}\hbar\partial_{x}=\widehat{p}_{x}-\hbar K_{\theta x},\\\widehat{q}_{y}&=-\dot{\iota}\hbar\partial_{y}=\widehat{p}_{y}-\hbar K_{\theta y}.
\end{align}
With this redefinition we have
\be
\widehat{H}=v\vec{\sigma}\cdot\left(\widehat{\vec{q}}-\vec{\delta}\right).
\ee
Comparing the above Hamiltonian with the Hamiltonian of massless Dirac electron in a magnetic field, we can identify that the shift in the Dirac point is equivalent to the application of vector potential $\vec{A}=-e\vec{\delta}/c$ produced by an external magnetic field \cite{Ghosh_2009} or associated with lattice defects/strain \cite{PhysRevB.73.125411}.
Translational symmetry along the $y$-direction suggests the following auxillary spinor as an ansatz
\be
\widetilde{\Psi}\left(\vec{r}\right) = \left[\begin{array}{c}
\widetilde{\psi}_{A}\left(x\right)\\
\widetilde{\psi}_{B}\left(x\right)
\end{array}\right]e^{\dot{\iota}q_{y}y}.
\ee
Following a calculation similar to that of Sec.\ \ref{sec:MDF}, we arrive at the following solution for $\widetilde{\psi}_{A}$ and $\widetilde{\psi}_{B}$:
\begin{align}
\widetilde{\psi}_{A}\left(x\right)&=a\,e^{\dot{\iota}qx}+b\,e^{-\dot{\iota}\left(q-2\delta_{x}\right)x},\\ 
\widetilde{\psi}_{B}\left(x\right) &= a\,e^{\dot{\iota}qx+\dot{\iota}\phi}-b\,e^{-\dot{\iota}\left(q-2\delta_{x}\right)x-\dot{\iota}\phi},
\label{eq: TBGAux}
\end{align}
where
\be
\sin\phi=\frac{\hbar v\left(q_{y}-\delta_{y}\right)}{E},
\ee
and $\phi$ represents the angle between the $y$-momenta and the $x$-momenta of the electron with respect to the local $K$ point. It is straightforward to show that $q$ satisfies the following energy-momentum relationship:
\be
E^{2}=\left(\hbar v\right)^{2}\left[\left(q-\delta_{x}\right)^{2}+\left(q_{y}-\delta_{y}\right)^{2}\right].
\ee

Now let us turn our attention into the case of a single domain wall located at $x=l$, where
\begin{align}
    v_{F}\left(x\right)&=v_{0}  \Theta\left(l-x\right)+v_{1} \Theta\left(x-l\right),\\\delta_{x}\left(x\right)&=\delta_{x1} \Theta\left(x-l\right),\\
    \delta_{y}\left(x\right)&=\delta_{y1} \Theta\left(x-l\right).
    \label{eq: TBGDWwall}
\end{align}
In the analogy with a magnetic vector potential, the shift $\vec{\delta}$ is equivalent to a sheet of magnetic field located at the domain wall and pointing perpendicular to the TBG, with magnitude
\begin{align}
\begin{split}
B_{z}&=\partial_{x}A_{y}-\partial_{y}A_{x},\\&=-\frac{e}{c}\delta_{y1}\delta\left(x-l\right).    
\end{split}
\end{align}
It is interesting to see that transitional symmetry along the $y$-direction implies that the shift of the Dirac point in the $x$-direction is not a physically relevant perturbation. Just as the vector potential can affect physical observables only through its curl $\vec{B} = \nabla \times A$, so only the shift in the Dirac point in the direction \emph{parallel} to the domain wall can enter the transmission coefficient.

Using the conservation of $q_{y}$, it is possible to relate $\sin\phi_{0}$ and $\sin\phi_{1}$:
\be
\sin\phi_{1}=\left(\frac{v_{1}}{v_{0}}\right)\sin\phi_{0}-\left(\frac{v_{1}}{v_{0}}\right)\frac{\hbar v_{0}\delta_{y1}}{E}.
\ee
Defining $u_{1}\equiv v_1/v_0$ and $\Delta_{1}\equiv \hbar v_{1}\delta_{y1}/E$, we have 
\be
\sin\phi_{1}=u_{1}\sin\phi_{0}-u_{1}\Delta_{1}.
\label{eq: NewSnell}
\ee
This equation represents a modified version of Snell's law for an interface separating two twist angle domains. Here, $\Delta_{1}$ is a dimensionless quantity, which is a proxy for how much the $K$ point of the MBZ shifts along the $y$-axis when passing across the domain wall. Due to the presence of this additional term $-u_1 \Delta_1$ in Snell's law, the symmetry between opposite incidences angles, $\phi_{0}$ and $-\phi_{0}$, is broken, and electrons with opposite incidences angles are refracted by different amounts. When $\Delta_1 \ll 1$, so that the shift in the Dirac point is small compared to the electron's momentum, one can expect near-Klein tunneling. As we show below, the leading effect of small $\Delta_{1}$ is to shift the angle of perfect transmission away from zero. On the other hand, if $\Delta_{1}\gg1$, the transmitted electron states are gapped and exist only as an evanescent modes. In this latter case, transmission through a single barrier is exponentially small in $D$.

By constructing the transfer matrix similar to the previous section (see Appendix \ref{sec: TBGTransMat}), one can find the reflection probability of a single domain wall to be
\be
R_{1}=\frac{1-\cos\left(\phi_{1}-\phi_{0}\right)}{1+\cos\left(\phi_{1}+\phi_{0}\right)},
\label{eq: newR1}
\ee
which is same as Eq.~(\ref{eq: T1}) except that the relation between $\phi_{0}$ and $\phi_{1}$ is modified [see Eq.~(\ref{eq: NewSnell})]. In the remainder of this section we will consider the case where $\Delta_1 \ll 1$, so that the electron momentum is relatively large. As we estimate below, in TBG this limit typically corresponds to electron energies $E \gtrsim 0.1$\,meV. In this case of small $\Delta_1$, one can make a Taylor expansion of Eq.~(\ref{eq: newR1}) around $\phi_{0}=0$ and $\Delta_{1}=0$, which gives
\be
R_{1}\approx\frac{u_{1}^{2}\Delta_{1}^{2}}{4}+\frac{1}{2}u_{1}\left(u_{1}-1\right)\Delta_{1}\phi_{0}+\frac{1}{4}\left(u_{1}-1\right)^{2}\phi_{0}^{2}.
\label{eq: R1TBG}
\ee
This equation suggests that perfect transmission happens at {$\phi_{0}=(u_{1}\Delta_{1})/(1-u_{1})$} rather than at normal incidence. Since this angle depends on the local Fermi velocity and the random vector potential, it varies across the sample. This situation is similar to that of the Brewster angle in optics, for which perfect transmission of polarized light across an interface occurs at an angle that depends on the local refractive indices. (Including higher-order terms in $\Delta_1$ in fact gives a finite reflection coefficient $R \sim u_{1}^{4}\Delta_{1}^{4}$ even at the optimal transmission angle.)

We can now calculate the statistical properties of the transmission through many domains of twist angle, as in the setup of Fig.\ \ref{fig:domains}. In addition to the random velocities $u_{i}$ associated with each domain, here we also account for the random shift of the Dirac point represented by the variables $\Delta_{i}$. Since in TBG the randomness in $u_{i}$ and $\Delta_{i}$ arise from the same variation in twist angle and are therefore not independent, we consider the situation
\be
\Delta_{i}=\lambda\epsilon_{i}\qquad u_{i}=1+\beta\epsilon_{i},
\ee
where $\lambda$ and $\beta$ are dimensionless experimental parameters.  (For a general situation where the Fermi velocity disorder and random energy shift are independent, refer to Appendix \ref{sec:IND}.) We assume a uniform distribution for the twist angle variation $\epsilon_{i}$:
\be
    \begin{split}
        P\left[\epsilon_{i}\right]=\left\{ \begin{array}{l r}
\frac{1}{\mu} & \text{if } -\frac{\mu}{2}\leq\epsilon_{i}\leq\frac{\mu}{2}\\
0 & \text{otherwise}
\end{array}\right..
    \end{split}
    \label{eq: TBGPDF}
\ee

Before proceeding to discuss the transmission, it is worth pausing to make an estimate of the typical numerical values of the parameters $\lambda, \beta$ and $\mu$. In TBG the first magic angle is $\sim 1.1^{\circ}$ and current experiments show a typical variation in twist angle of $\sim 10\%$ \cite{Uri_2020}, which gives $\mu\approx0.004$. Within the continuum description used here \cite{PhysRevB.84.045436}, $\beta = \cot(\theta/2) / 2$ [see Eq.\ (\ref{eq: newvf})], so that using $\theta\sim1.1^{\circ}$ gives $\beta\approx50$. Thus, in typical TBG samples near the magic angle, the dimensionless random energy shift $\Delta_{i}\ll1$ for all electron energies above $0.1$\,meV, as mentioned above. For example, if $\Delta_{i}\sim0.01-0.1$, then $\lambda$ is in the range $1$ -- $25$ for $\mu=0.004$.

In order to calculate the inverse localization length, $D/\xi \approx \left\langle R_{i} \right\rangle$, we first Taylor expand $R_{i}$ in terms of $\epsilon_{i}$, $\epsilon_{i+1}$ and $\phi_{0}$ and then proceed to take the expectation value over the random twist angle. This calculation gives
\be
\langle R_{i}\rangle\approx\frac{\mu^{2}\left(2\lambda^{2}+\beta^{2}\right)}{24}\left(\phi_{0}-\phi_{\text{min}}\right)^{2}+R_{\text{min}},
\label{eq: TBGDWRi}
\ee
where 
\be
\phi_{\text{min}}=\frac{\lambda\beta}{2\lambda^{2}+\beta^{2}}\quad \text{ and}\quad R_{\text{min}}=\frac{\mu^{2}\lambda^{2}}{24}\left[1-\frac{\beta^{2}}{2\lambda^{2}+\beta^{2}}\right].
\ee
The finite reflection probability $R_\text{min} > 0$ arises because the Brewster-like angle associated with transmission varies from one interface to another, and therefore on average there is no Klein tunneling. Instead, the disorder average gives a minimum of reflection at a nonzero angle $\phi_{\text{min}}$. Using the experimental parameters presented earlier in this section, we get an estimate of $R_\text{min}$ that is extremely small, in the range $10^{-6}-10^{-9}$.  This small value of $R_\text{min}$ implies a very long localization length for electrons incident at the angle of minimum reflection, $\phi_{\text{min}}$. Correspondingly, the collimation of a dispersed electron beam is centered around a value $\phi_{min}$ that depends on the electron energy $E \propto 1/\lambda$ [see Fig.~\ref{fig:mainresult}(b)]. This result implies that twist angle disorder may be used not only to collimate electrons but also to filter them in energy, since the direction of the transmitted electron beam depends on its energy.

\begin{figure}[htb]
\centering
\includegraphics[width=1.0 \columnwidth]{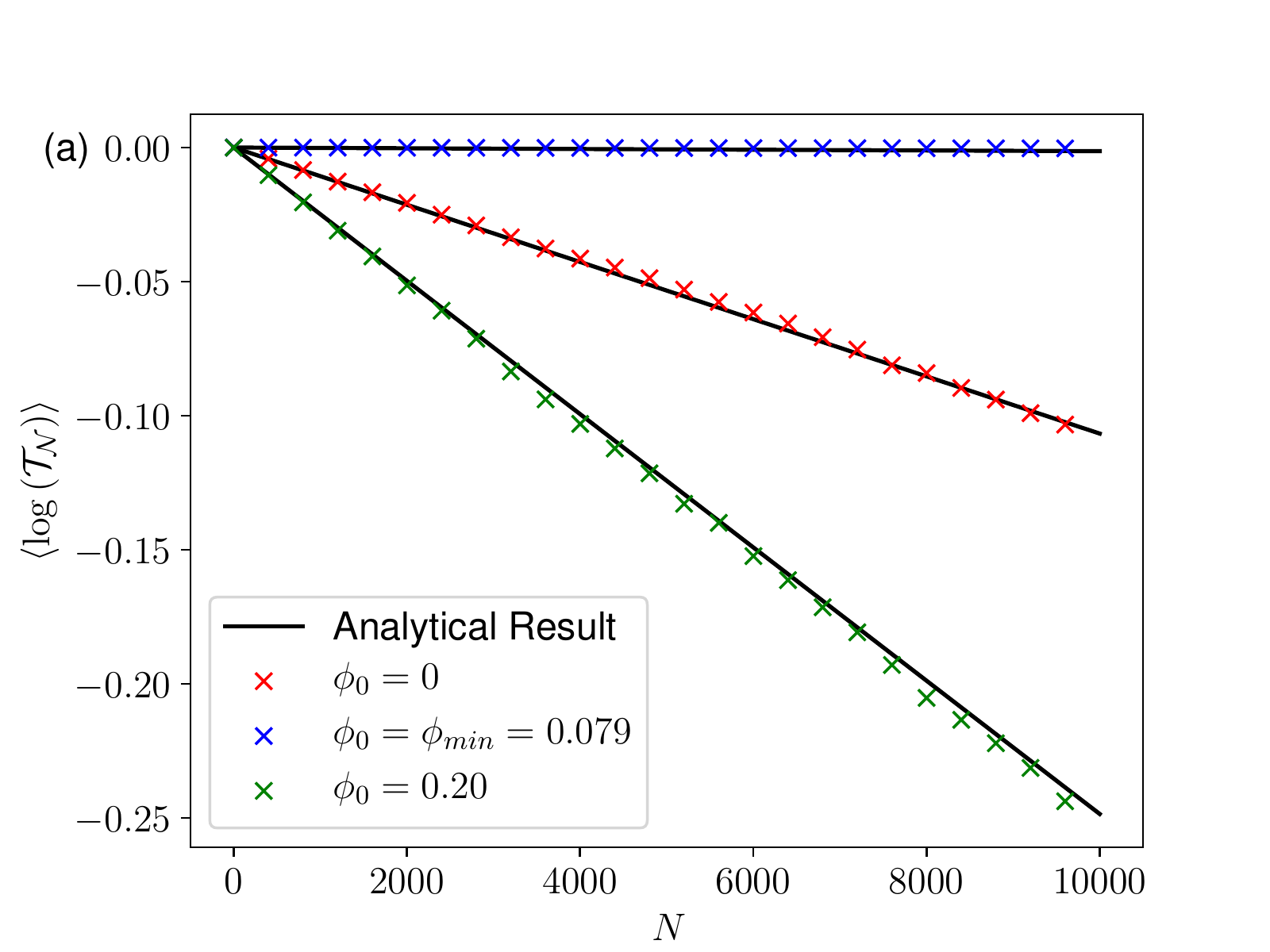}
\includegraphics[width=1.0 \columnwidth]{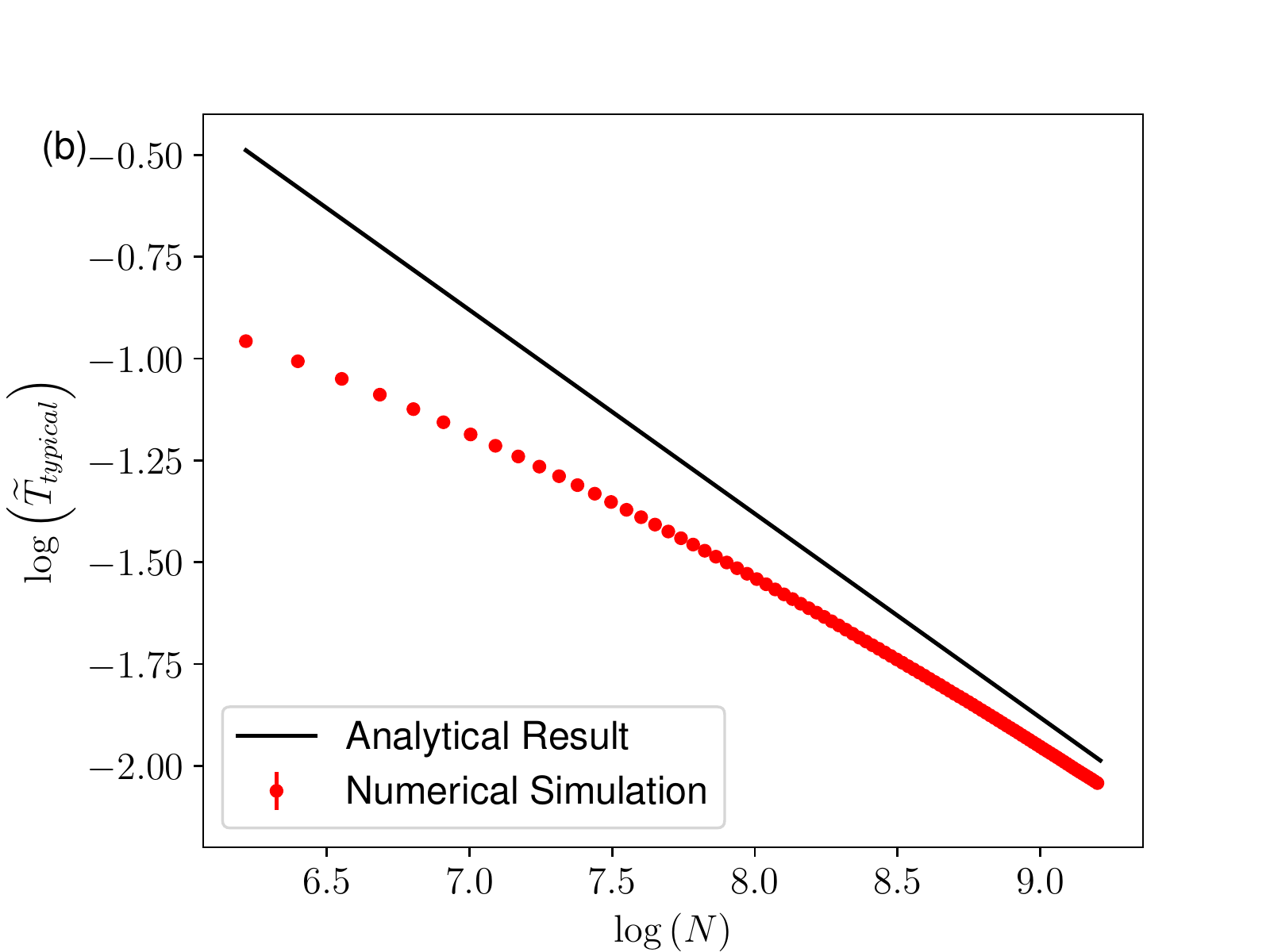}
\caption{Decay of the transmission amplitude with the number of interfaces for a minimal model of TBG, for which variations in twist angle produce both a change in velocity and a shift of the Dirac point in reciprocal space. (a) The average of $\log\left(\mathcal{T}_N\right)$ is plotted as a function of $N$ for three different initial incidence angles: $\phi_{0}=0, \,\phi_{\text{min}}\text{, and } 0.20$. The localization length increases when $\phi_{0}$ increases from $0$, reaches a maximum at $\phi_{0}=\phi_{\text{min}}$ and then decreases again. (b) The average of the typical transmission over incidence angle is plotted as a function of $\log\left(N\right)$. The analytical curve is plotted using Eq.\ (\ref{eq: TBGTypAvg}), which corresponds to the $N \rightarrow \infty$ limit, and the numerical data is generated using the transfer matrix in Eq.\ (\ref{eq: TransMatTBG}). The parameters used for the simulation are: $\mu=0.004$, $\lambda=4$ and $\beta=50$.}
\label{fig:dom}
\end{figure}

As in the previous section, the value of the transmission $\mathcal{T}_{N}$ is log-normal distributed across an ensemble of samples [see Eq.~(\ref{eq: lognormal})], with the mean value of $\log\left(\mathcal{T}_N\right)$ given by $N$ times Eq.~(\ref{eq: TBGDWRi}). The analytical prediction for the mean of $\log\left(\mathcal{T}_N\right)$ is plotted in Fig.~\ref{fig:dom}(a), along with numerical results obtained using the transfer matrix approach. Using Eq.~(\ref{eq: TBGDWRi}), we can calculate the angle-averaged typical transmission [see also Eq.~(\ref{eq: typicaldef})], which gives
\be
\widetilde{T}_{\text{typical}}=\sqrt{\frac{24}{\pi\mu^{2}\left(2\lambda^{2}+\beta^{2}\right)}}\frac{\exp{\left(-NR_{\text{min}}\right)}}{\sqrt{N}}.
\label{eq: TBGTypAvg}
\ee
This expression is plotted in the Fig.\ \ref{fig:dom}(b) along with numerical results.

\section{Conductance and Fano Factor}
\label{sec: condfano}

In the preceding sections we calculated the typical transmission coefficient $T_\text{typical}(\phi_0)$ for electrons as a function of the initial incident angle $\phi_0$, considering both Fermi velocity disorder and a more general model of twist angle disorder in TBG. In this section we briefly discuss the implications of these results for the electrical conductance and the Fano factor.

According to the Landauer formula, the electrical conductance of a sample with width $W$ is given by \cite{PhysRevB.84.195404}
\be
G=G_{0}\int_{-\frac{\pi}{2}}^{\frac{\pi}{2}}d\phi_{0}\;\cos{\phi_{0}}\;T_{\text{typical}}\left(\phi_{0}\right),
\label{eq: cond}
\ee
where $G_{0}$ is given by the ballistic conductance $\left(\frac{4e^{2}}{h}\right)\times\text{W}$. For the small angles $\phi_0$ that dominate $T_\text{typical}$ at large $N$, $\cos{\phi_{0}}\approx1$. Hence, the conductance is effectively proportional to the angle-averaged transmission coefficient $\widetilde{T}_{\text{typical}}$, and $G$ inherits the same power-law decay [see Eq.\ (\ref{eq: angavgttyp})], $G \propto 1/\sqrt{N}$. One can also calculate the conductivity of the sample, given by
\be
\sigma=\left(\frac{ND}{W}\right)G.
\ee
Thus, $\sigma\sim\sqrt{N}$. 
A similar growth of the conductivity with the square root of system size was predicted by Ref.\ \onlinecite{Titov_2007} for the case of massless Dirac electrons with uncorrelated, one-dimensional scalar disorder potential. 

Another experimentally relevant quantity is the Fano factor $F$ associated with the shot noise, defined by
\be
F=1-\frac{\int_{-\frac{\pi}{2}}^{\frac{\pi}{2}}d\phi_{0}\;\cos{\phi_{0}}\;{T}_{\text{typical}}^{2}\left(\phi_{0}\right)}{\int_{-\frac{\pi}{2}}^{\frac{\pi}{2}}d\phi_{0}\;\cos{\phi_{0}}\;{T}_{\text{typical}}\left(\phi_{0}\right)}.
\ee
For large $N$, where the transmission is similarly dominated by small incident angles $\phi_0$, we arrive at
\be
F = 1 - \frac{1}{\sqrt{2}}\approx 0.29.
\ee
This same value of the Fano factor has been derived for a smooth ballistic $p-n$ junction  in monolayer graphene by Cheianov et al.\ \cite{PhysRevB.74.041403}, where it arises from a similar dependence of the transmission on the incident angle.

\section{Summary and conclusion}
\label{sec:conclusion}

In this paper, inspired by the recent experimental observation of random domains of twist angle in TBG, we have considered the effect of Fermi velocity disorder on the transport of massless Dirac fermions. We have focused in particular on the model of quasi-one-dimensional disorder depicted in Fig.\ \ref{fig:domains}, which has a close analogy with the physics of a ``transparent mirror'' in optics \cite{Berry_1997}, which itself can be mapped onto Anderson localization. Unlike the optical problem, however, the localization length in our problem diverges as the angle of incidence approaches zero [Eq.\ \ref{eq: xi}], and for such angles the material is transparent due to Klein tunnelling. Since the inverse localization length has a quadratic dependence on the incidence angle up to leading order, the angle-averaged typical transmission of the system decays with the system size in a power-law manner: $\widetilde{T}_{\text{typical}}\propto1/\sqrt{N}$ [Eq. (\ref{eq: angavgttyp})]. These results have a direct relationship to the conductance of the sample, which follows the same power law behavior [Eq.\ (\ref{eq: cond})]. The corresponding Fano factor is $1-1/\sqrt{2}$.

Whether these results apply directly to twist angle disorder in TBG depends, in principle, on the details of how the twist angle varies from one domain to another. If one of the two layers is perfectly fixed while the other is allowed to twist in a spatially random manner, then Klein tunneling persists for only of the two valleys, leading to a disorder-induced valley filtration in which only one of the two valleys is transmitted at normal incidence. On the other hand, in the more realistic case where both layers undergo random twisting, one must consider the spatial modulation of both the  Fermi velocity and a random gauge field that describes the shifting of Dirac points in momentum space. This gauge field leads to a modification of Snell's law [Eq.\ (\ref{eq: NewSnell})] and a near-perfect transmission of electrons at a non-zero incidence angle, similar to the Brewster angle in optics. Since the Brewster-like angle varies from one domain wall to another, the overall reflection probability has a nonzero minimum [Eq.\ (\ref{eq: TBGDWRi})], which occurs at non-zero incidence. Making numerical estimates based on recent TBG experiments, we found this minimum reflection coefficient to be very small, which suggests essentially perfect transmission at a nonzero angle. Thus, TBG may also perform disorder-induced collimation of incoming electrons, but to a nontrivial angle that depends on the energy of electrons.

It is also worth pointing out that the modified Snell's law in TBG [Eq.\ (\ref{eq: NewSnell})] implies that for suitable values of $u_{1}$ and $\Delta_{1}$, it is possible to have negative refraction.  This situation gives rise to Veselago lensing, as in p-n junctions in monolayer graphene \cite{Cheianov_2007}. 


Of course, the model considered in this work is only a highly simplified case, for which the disorder is one-dimensional. The corresponding translational symmetry in the $y$-direction has allowed us to use a simple transfer matrix approach and to map the problem onto one-dimensional Anderson localization.  A more realistic model with random two-dimensional domains does not admit a similarly simple transfer matrix description.  Still, the results presented here, in particular the phenomena of disorder-induced valley filtration and collimation, may provide inspiration for new directions in Dirac fermion optics based on tailored twist patterns.

\acknowledgments
We are thankful to C. W. Beenakker, J.\ Bertolotti, J. H. Wilson, and Cyprian Lewandowski for useful discussions. Computations were performed using the Unity cluster at The Ohio State University.
\bibliographystyle{apsrev4-1}
\bibliography{Dirac_Random_Velocity}

\widetext 
\appendix
\section{Statistical properties of the transmission probability: Mapping to Fokker-Planck equation}
\label{sec: fokk}

\begin{figure}
\centering
\includegraphics[scale=0.75]{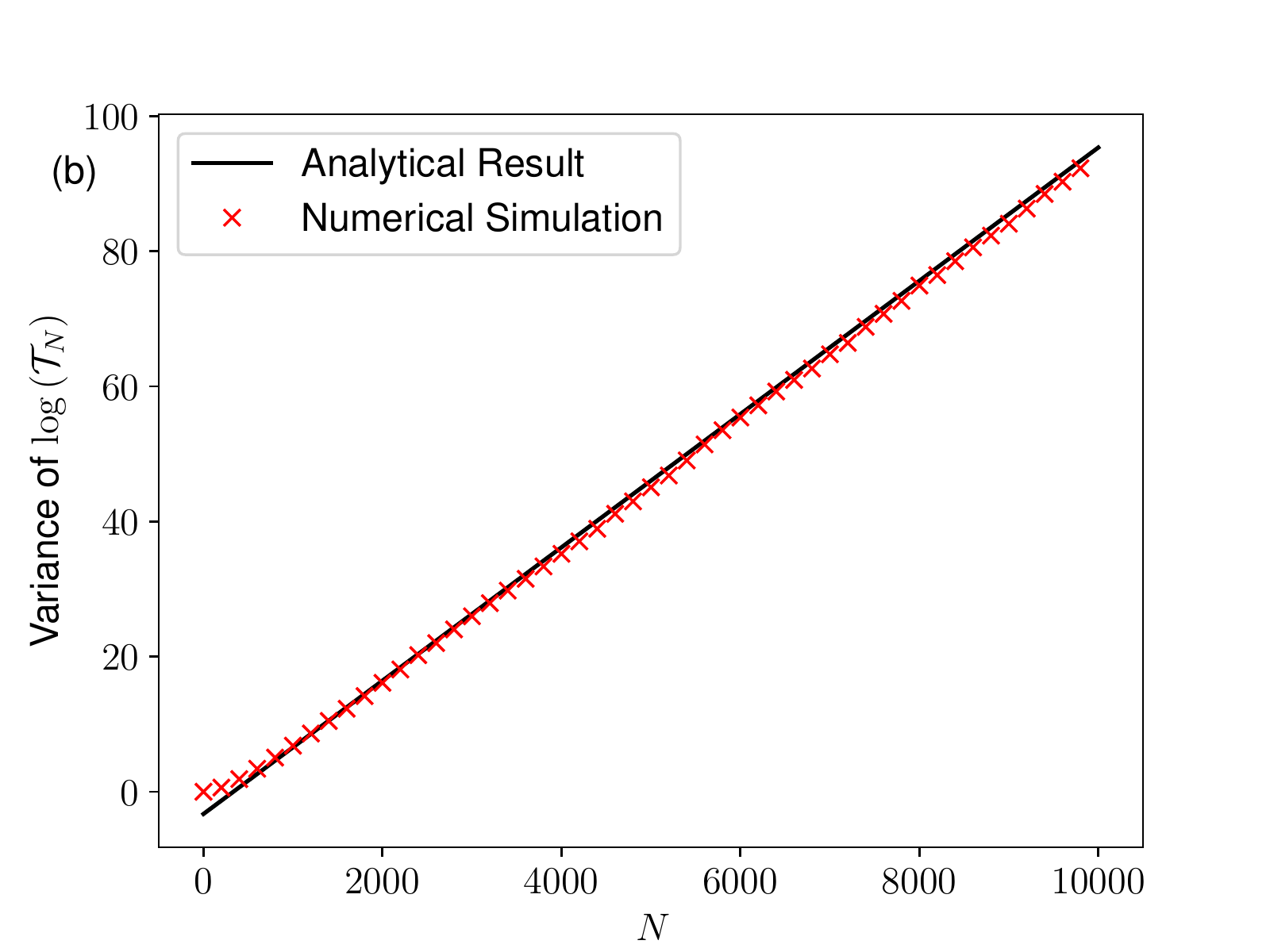}
\caption{The variance of $\log\left(\mathcal{T}_{N}\right)$ is plotted as a function of $N$ for an incidence angle of $\phi_{0}=\pi/10$, and width of the velocity distribution $w=1$. The solid black line represent the analytic prediction as in Eq.~(\ref{eq: varLogT}) and the crosses represents the numerical data obtained using the transfer matrix approach.}
\label{fig: varlog}
\end{figure}

Here we provide more details about the statistical properties of the transmission probability. In particular, we can obtain the variance of $\log\left(\mathcal{T}_{N}\right)$. If we define $\Delta\log\left(\mathcal{T_{N}}\right)$ as $\log\left(\frac{\mathcal{T}_{N+1}}{\mathcal{T}_{N}}\right)$, then by Eq.~(\ref{eq: T12})
\be
\Delta\log\left(\mathcal{T_{N}}\right)=\log T_{1}-2\log\left|1+\sqrt{R_{1}R_{N}}e^{\dot{\iota}\Phi_{NN+1}}\right|,
\ee
where $\Phi_{NN+1}$ represents the phase accumulated during one complete internal reflection path between interface $N$ and interface $N+1$.
Thus, for small incidence angle, $R_{1}\ll1$, we have
\be
\Delta\log\left(\mathcal{T_{N}}\right)\approx-R_{1}-2\sqrt{R_{1}R_{N}}\cos\Phi_{NN+1}+R_{1}R_{N}\cos2\Phi_{NN+1}
\ee
to leading order in $R_{1}$.
Averaging this expression over random velocities, and employing the same random phase approximation used in Eq.~(\ref{eq: logT12}), we arrive at
\be
\left\langle \Delta\log\left(\mathcal{T_{N}}\right)\right\rangle \approx -\left\langle R_{1}\right\rangle. 
\label{eq: DelLog}
\ee
This is same as the result obtained in Eq.~(\ref{eq: logTN}). The variance of $ \Delta\log\left(\mathcal{T_{N}}\right)$ can be found to be
\be
\mathrm{var}\left[\Delta\log\left(\mathcal{T_{N}}\right)\right] \approx 2\left\langle R_{1}\right\rangle.
\label{eq: varDelLog}
\ee
This expression implies that the variance in transmission is dominated by the randomness in the phase $\Phi_{NN+1}$ across a domain, rather than in the values of $T_N$ or $T_1$.  For $N$ interfaces, the corresponding variance in the transmission is
\be
\mathrm{var}\left[\log\left(\mathcal{T_{N}}\right)\right] \approx 2N\left\langle R_{1}\right\rangle.
\label{eq: varLogT}
\ee
A more elaborate calculation can be used to show that there is a correction of $-\pi^{2}/3$ to the variance in Eq.~(\ref{eq: varLogT}) \cite{ABRIKOSOV1981997}. In Fig.\ \ref{fig: varlog} we plot the variance calculated from numerical simulation along with this analytical prediction.
Equation (\ref{eq: varLogT}) also establishes that the relative fluctuation in $\log\left(\mathcal{T}_{N}\right)$ is
\be
\frac{\text{var}\log\left(\mathcal{T}_{N}\right)}{\left\langle \log\left(\mathcal{T}_{N}\right)\right\rangle ^{2}}\propto\frac{1}{N}.
\ee
This decay of the relative fluctuation as the number of interfaces grows implies self-averaging, and confirms that $\log\left(\mathcal{T}_{N}\right)$ is normally distributed in the limit of sufficiently large $N$. However, since $\mathcal{T}_{N}$ is log-normal distributed, $\mathcal{T}_{N}$ itself is not self-averaging and the difference between $\left\langle \mathcal{T}_{N}\right\rangle$ and $e^{\left\langle \log\left(\mathcal{T}_{N}\right)\right\rangle }$ becomes exponentially different at large $N$. One can extract the distribution for $\mathcal{T}_{N}$; since the derivation is cumbersome, here we only present the results. We refer interested readers to the calculations in the lecture notes by M\"{u}ller and Delande \cite{mller2010disorder}. If we define $\Delta\mathcal{T}_{N}=\mathcal{T}_{N+1}-\mathcal{T}_{N}$, then to leading order in $\mathcal{R}_1$
\be
     \Delta\mathcal{T}_{N} \approx-\mathcal{T}_{N}\left(2\sqrt{\mathcal{R}_{1}\mathcal{R}_{N}}\cos\Phi+\left(\mathcal{R}_{N}+1-4\mathcal{R}_{N}^{2}\cos^{2}\Phi\right)\mathcal{R}_{1}\right).
\ee
Following the previous averaging process we obtain
\begin{align}
\left\langle \Delta\mathcal{T}_{N}\right\rangle &=-\left\langle R_{1}\right\rangle \left\langle \mathcal{T}_{N}^{2}\right\rangle ,\label{eq: moments1}\\\left\langle \left(\Delta\mathcal{T}_{N}\right)^{2}\right\rangle &=\left\langle R_{1}\right\rangle \left\langle 2\mathcal{T}_{N}^{2}\left(1-\mathcal{T}_{N}\right)\right\rangle . 
\label{eq: moments2}
\end{align}
All higher order moments of $\Delta\mathcal{T}_{N}$ vanish. Equations (\ref{eq: moments1}) and (\ref{eq: moments2}) can be used to map the probability distribution of $\mathcal{T}_{N}$ to a Fokker-Planck equation, and the result is obtained as
\be
P\left(\log\mathcal{T}_{N}\right)=\frac{1}{2\sqrt{\pi N\left\langle R_{1}\right\rangle }}\exp \left( - \frac{\left(\log T+N\left\langle R_{1}\right\rangle \right)^{2}}{4N\left\langle R_{1}\right\rangle }\right).
\ee
A closed form expression for $P\left(\mathcal{T}_{N}\right)$ is not possible, though in the limit $N\left\langle R_{1}\right\rangle \gg1$ all the moments can be calculated as a closed expression,
\be
\left\langle \mathcal{T}_{N}^{n}\right\rangle =\frac{\pi^{\frac{3}{2}}}{2}\left[\frac{\Gamma\left(n-\frac{1}{2}\right)}{\Gamma\left(n\right)}\right]^{2}\left(N\left\langle R_{1}\right\rangle \right)^{-\frac{3}{2}}\exp\left(-\frac{N\left\langle R_{1}\right\rangle }{4}\right).
\ee
In particular, we can see that the average fluctuation in $\mathcal{T}_{N}$ grows exponentially with $N$, which reflects the failure of self-averaging. 
\section{Transfer Matrix for TBG with twist disorder}
\label{sec: TBGTransMat}
The transfer matrix for an interface separating two twist angle domains [see Eq.~(\ref{eq: TBGDWwall})] can be calculated using the using continuity of the auxiliary spinor given in Eq.~(\ref{eq: TBGAux}) . This process gives
\be
M=\frac{1}{2\sqrt{\cos\phi_{1}\cos\phi_{2}}}\left[\begin{array}{cc}
e^{\dot{\iota}\left(q_{1}-q_{2}\right)l}\left(e^{\dot{\iota}\phi_{1}}+e^{-\dot{\iota}\phi_{2}}\right) & e^{-\dot{\iota}\left(q_{1}+q_{2}\right)l}\left(e^{-\dot{\iota}\phi_{2}}-e^{-\dot{\iota}\phi_{1}}\right)\\
e^{\dot{\iota}\left(q_{1}+q_{2}-2\delta_{x1}\right)l}\left(e^{\dot{\iota}\phi_{2}}-e^{\dot{\iota}\phi_{1}}\right) & e^{-\dot{\iota}\left(q_{1}-q_{2}+2\delta_{x1}\right)l}\left(e^{-\dot{\iota}\phi_{1}}+e^{\dot{\iota}\phi_{2}}\right)
\end{array}\right]
\label{eq: TransMatTBG}
\ee

\section{Disorder-averaged transmission coefficient when the Fermi velocity and random energy shifts are independent}
\label{sec:IND}
Here, we consider the case where the relative change in velocity $u_{i}$ and the relative momentum shift $\Delta_{i}$ of the Dirac point across a domain boundary are taken to be independent random variables. This is not the case of TBG with twist angle disorder, for which both variables are coupled to the twist angle. For the value of $u_{i}$ we use the same uniform probability distribution as before [see Eq.~(\ref{eq: PDF})], while for the value of  $\Delta_{i}$ we use the following distribution:
\be
P\left[\Delta_{i}\right]=\left\{ \begin{array}{c}
1/\alpha\qquad\text{if}\qquad-\frac{\alpha}{2}\le u\le\frac{\alpha}{2}\\
0\qquad\text{otherwise}
\end{array}\right..
\ee
Using the above disorder distribution we will obtain 
\be
\left\langle R_{i}\right\rangle \approx \frac{w^{2}\phi_{0}^{2}}{24}+\frac{\left(12+w^{2}\right)\alpha^{2}}{288},
\label{eq: avgRiInd}
\ee
where $\langle.\rangle$ denotes the average over values of both $u_i$ and $\Delta_i$.

We can see that, when the energy shift $\Delta_i$ is taken to be independent of the change in velocity, symmetry between opposite reflection angles $ \pm \phi_{0}$ is restored. Equation \ref{eq: avgRiInd} suggests that the localization length increases as $\phi_{0}$ decreases and reaches a finite value at $\phi_{0} = 0$. We can show that angle averaged typical transmission in this case would be
\be
\widetilde{T}_{typical} = \sqrt{\frac{24}{\pi w^{2}}}\left(\frac{\exp[-N\left(\frac{\left(12+w^{2}\right)\alpha^{2}}{288}\right)]}{\sqrt{N}}\right).
\ee

\end{document}